# Synergetic Application, Equations on Rule of Law and Two-Party Mechanism

*Yi-Fang Chang*
*Department of Physics, Yunnan University, Kunming 650091, China*
(E-mail: yifangchang1030@hotmail.com)

**Abstract**

Based on the synergetic equations and its application, we propose the equations on the rule of law. From these equations we may prove mathematically that a society of the rule of law cannot lack any aspect for three types of the legislation, the administration and the judicature. Otherwise, we propose an equation of corruption, and discuss quantitatively some threshold values for a system into corruption. Moreover, from synergetics we obtain the Lorenz model, which may be a visualized two-party mechanism as a type of stable structure in democracy. A developed direction of society should be the combination from macroscopic to microscopic order, from an actual capable handling to an ideal pursuance.
Key words: nonlinear equations, rule of law, synergetics, democracy.
PACS number(s): 05.45.-a; 89.65.-s; 01.75.+m.

**I.Introduction**

As early as old Greece Plato has discussed various structures on the division of work as well as cooperation in an ideal society. According to the social contractual theory, the government and nation are created to pass a social contract. First, Thomas of Aquino made statement for this theory [1]. He discussed that the person's life demands division of work, and the society must unite, such the purpose of a nation should be to seek together happiness of the society. Later Thomas Hobbes and John Locke, et al., developed this theory. Locke stipulated that duty of government should be to protect the basic right of people, so the sovereignty is similar to other persons and is subjected to the check and supervision by the contractual contents. People are entitled to overset government, if which breached contract. Jean Rousseau expatiated sufficiently the social contractual theory [2]. The laws should be the outcome of people's majority opinions, and the social common contract must agree with collectivity. Only this type of society can ensure together freewill cooperation of freeman. Otherwise, any law is only a paper of the ruler order.

Social contract theory provides the rationale behind the historically important notion that legitimate government must be derived from the consent of the governed. Social contract implies that the people give up some rights to a government in order to receive or jointly preserve social order. It is in an individual"˜s rational self-interest to voluntarily subjugate the freedom of action one has under the natural right in order to obtain the benefits provided by the formation of social structures. The social contract is a basic of all civilization.

On the most extensive meaning, Montesquieu proposed that any law is a necessity relation. This is an expansion of the natural law. According to Montesquieu's principle, political system classifies democracy, monarchy and despotism. He advocated constitutionalism and the separation of powers, the preservation of civil liberties and the rule of law, and the idea that political and legal institutions ought to reflect the social and geographical character of each particular community [3]. Later, the modern theory of law is expounded completely, in which a basic way is the separation of powers, and the three powers of legislation, judicature and administration cooperate and be restricted each other. Now people think, the power self tends always to autarchy and corruption, therefore, a good society must restrict any power by other power. W.Morrison discussed systematically the jurisprudence from the Greeks to post-modernism [4].

## 2.The equations on the rule of law derived from synergetics

Synergetics proposed by H.Hanke is an interdisciplinary field [5,6]. It deals with systems that are composed of many individual parts and that are able to produce spatial, temporal or functional structures by means of self-organization. Its central issue is the emergence of new qualities at macroscopic scales. Synergetics has developed some specific vocabulary and basic mathematical methods, for example, the order parameters and the slaving principle, etc. It as a quantitative theory is applied to very wide regions. The social contract is namely a special form of the social synergetics [7].

We combine the modern theory of law and synergetics, and research the basis of the rule of law by some mathematical methods. First, we apply the basic equations of synergetics [5]:

$$
\begin{aligned}
db/dt &= -kb - i\sum g_\mu \alpha_\mu + F(t), \\
d\alpha_\mu/dt &= -\gamma\alpha_\mu + ig_\mu{}^* b\sigma_\mu + \Gamma_\mu(t), \\
d\sigma_\mu/dt &= \gamma_{11}(d_0 - \sigma_\mu) + 2i(g_\mu \alpha_\mu b^* - c.c.) + \Gamma_{\sigma,\mu}(t),
\end{aligned} \tag{1}
$$

where b is the amplitude of field, $\alpha_\mu$ is the atomic dipole moment, and $\sigma_\mu$ is the atomic inversion. When k<0, these equations may describe the productions of laser and other order systems. From these equations we discuss mathematically the establishment of the rule of law and corresponding evolvement.

First, we define various quantities: b is a social field, which includes three aspects: law, strength of executed law (equity, strictness) and supervision, including one by people, society and consensus, etc. $\alpha_\mu$ may be the attitude of leader. $\sigma_\mu$ as the atomic inversion may correspond to human number or degree of keeping law. In this case three equations represent respectively: 1.Change of law, which relates to level and purpose of a nation, and resistances, etc. 2.Change of the attitude of leader and their level, etc. 3.Change of degree of keeping law, which relates to many complex social factors, for example, the numbers of concourse and of bureaucracy, that is more, change of law is more. Moreover, these have always chanciness and fluctuation.

In the above-mentioned equations, if $b=0$, it corresponds to no law, hence the equations change into:

$$
\begin{aligned}
d\alpha_\mu/dt &= -\gamma\alpha_\mu + \Gamma_\mu(t), \\
d\sigma_\mu/dt &= \gamma_{11}(d_0 - \sigma_\mu) + \Gamma_{\sigma,\mu}(t).
\end{aligned} \tag{2}
$$

These are two independent line equations each other, their solutions are:

$$
\begin{aligned}
\alpha_\mu &= C\exp(-\gamma t) + \Gamma_\mu(t)/\gamma, \\
\sigma_\mu &= C'\exp(-\gamma_{11} t) + d_0 + \Gamma_{\sigma,\mu}(t)/\gamma_{11}.
\end{aligned} \tag{3}
$$

The Eqs.(3) represent the degree of keeping law (namely, the social morality) will decrease exponentially. Finally, $\alpha_\mu = \Gamma_\mu / \gamma$ for Eq.(3a), which is consistent with $\alpha_\mu = aF(t)$. This determines only chanciness. It shows that persons in power treat all at individual inclinations. Conversely, common people are treated unbendingly by these leaders. This is a typical autarchy government of completely lawless. When $t \to \infty$ for Eq.(3b), $\sigma_\mu = d_0$ represents that only a few with higher diathesis exists, which include a few autonomic functionary.

Using the same method, if $\alpha_\mu = 0$ in Eqs.(1), it corresponds to that various laws are only a nomenclature for power, the equations changes into symmetrical forms:

$$db/dt = -kb + F(t),$$
$$d\sigma_\mu /dt = \gamma_{11}(d_0 - \sigma_\mu) + \Gamma_{\sigma,\mu}(t), \quad (4)$$

These are also two independent line equations each other, their solutions have the same forms:

$$b = Ce^{-kt} + F(t)/k, \quad (5)$$

and (3b). The Eq.(5) represents the number of keeping law will decrease exponentially. Finally, $b = F(t)/k$, which is consistent with $b = -c\Gamma_\mu(t)$. This determines also only chanciness. It shows that although the laws exist, but persons of power handle laws at individual inclinations. Therefore, laws exist in name only, and all is still zero. For Eq.(3b), the solutions and conclusions are completely the same.

Thirdly, when $\sigma_\mu = 0$ in Eqs.(1), these equations change into

$$db/dt = -kb - i\sum g_\mu \alpha_\mu + F(t),$$
$$d\alpha_\mu /dt = -\gamma \alpha_\mu + \Gamma_\mu(t), \quad (6)$$

and $\alpha_\mu b^* = -d\Gamma_{\sigma,\mu}(t)$. The solution of (6b)

$$\alpha_\mu = Ce^{-\gamma t} + \Gamma_\mu(t)/\gamma, \quad (7)$$

replaces into (6a), then equation changes to linear one, and the above conclusions are the same.

These results show mathematically that the three powers: lawmaking, executed law and supervision, cannot be short of any one. If not, these equations will change into the linear equations, which cannot be restricted each other. Their solutions point unequivocally out: The just laws will be nothing left along with time. This shows also the restricted relations among legislation, administration and judicature. Contrarily, all from tyrant and fatuous ruler to ruffians is unruliness.

When synergetics is applied to society, the order parameters are the laws and their contents, the strength of executed law, and supervision. For any country the most essential law is a constitution. While individual power, morality and mass movement, etc., are only rapid relaxation parameter.

By the adiabatic approximation in synergetics, Eqs.(1) can change to the two simplifying equations:

$$db/dt = -kb + (g^2/\gamma)b\sum\sigma_\mu + F(t),$$
$$d\sigma_\mu/dt = \gamma_{11}(d_0 - \sigma_\mu) - (4g^2/\gamma)b^*b\sigma_\mu. \tag{8}$$

Further, theses equations may become a nonlinear equation, for example, a laser equation.

Usually, law likes a sword. If we describe rule of law by mathematical and physical method, the simplest equation of rule of law will be analogous to laser equation:

$$\frac{db}{dt} + \alpha b + \beta b^3 = F(t). \tag{9}$$

This determines the strength of executed law and human number on keeping law. In this case, the order parameter is $\alpha = C_0 - kD$. When D is smaller, and $D < C_0/k$, so $\alpha > 0$, and the potential is symmetrical and unstable for equation of rule of law. This is maintained mainly by some random processes F(t). When D increase to $D = C_0/k$, and $\alpha = 0$, then the phase transformation appears. When $D > C_0/k$ and $\alpha < 0$, the potential forms a metastable state broken symmetry. In this case, new phase on rule of law like laser will appear. In a social system, human number of keeping law achieves to a certainty amount, and degree of keeping law is stronger, then the slaving principle will show huge function, and the society of rule of law will be built. As Hayek said [8], law like morality, religion, language, etc., will form a spontaneous social order, i.e., a self-organization. This will overbear effectively those unlawful actions like a laser sword. It will be similar laser with high brightness, directionality and homogeneous light, and possess following characters: clarity (known law), directionality (keeping law) and oneness (everyone equality face law). Such men will become a type of rule-following animals [9].

## 3. A corruption equation

When number of keeping law decreases to a certain threshold value, and $D = C_0/k$, the equation (9) will become into:

$$\frac{dS}{dt} = 1 - \beta S^2, \tag{10}$$

its solution is:

$$S = cth(\beta t)/\sqrt{\beta}. \tag{11}$$

The social rule of law will decrease along with time, and finally, it will not exist. In this case, Eq.(10) becomes a difference equation:

$$X_{n+1} = 1 - \lambda X_n^2. \tag{12}$$

Assume that the replacement may represent corruptions from one level to another level, or continuous expansive corruption, this equation may be called to a corruption equation. For this, if corruption cannot be controlled effectively, the bifurcation with two periods will appear, and finally, the solution will achieve chaos, namely, whole social system will decompose and become confusion. From quantitative value of bifurcation-chaos, we may estimate that if the corruption members in any system or a certain level attain to 37.5%, a dangerous double periodic bifurcation will appear. If the members attain to 0.70057759...=70%, this system as whole will show an incurable dark. It is represented visually in a known figure from bifurcation to chaos.

From a point of view of law, the supervisions each other are a promoted cooperation. Corresponding self-discipline is only an autocriticism in an isolated system. This may bring that those noble people preserve one's moral integrity. But, faced a bad force colluded each other, only the self-discipline is invalid for the completion of a society. Therefore, excessively emphasized consciousness but no restriction with laws is adverse for country or nation. The absence of rule of law is out of a fundamental welfare for total country.

## 4. Lorenz model and two-party mechanism

Moreover, the Lorenz model may be obtained from the basic equations of synergetics (1) [10]. When we neglect various random effects $F(t), \Gamma_\mu(t), \Gamma_{\sigma,\mu}(t)$, and let $b \to x, \alpha_\mu \to y$ and $\sigma_\mu \to \sigma$, so Eqs.(1) are simplified to:

$$db/dt = dx/dt = -kx - ay, \tag{13}$$

$$d\alpha_\mu/dt = dy/dt = -by + cx\sigma, \tag{14}$$

$$d\sigma/dt = A - e\sigma + fxy. \tag{15}$$

Let $z = A - e\sigma$, so Eqs.(13)(14)(15) become the equations of the Lorenz model [11]:

$$dx/dt = -vx + ky, \tag{16}$$

$$dy/dt = ax - by - xz, \tag{17}$$

$$dz/dt = -cz + xy. \tag{18}$$

This result is a space mode. Usually, we suppose that all parameters are positive. If various parameters in Eqs.(16)(17)(18) take suitable values, a beautiful Lorenz strange attractor will be obtained. The social meaning of the Lorenz model is a visualized two-party mechanism. It shows that when the social-field strength as an order parameter achieves a certain threshold, a type of stable self-organized structure in democracy can be formed from chaos and synergetics.

An ideal men in society for the Chinese traditional culture should cultivate one's and family's noble morality, later manage country, even world. In a society of the Great Harmony, the material is very abundant and every personal consciousness is very high. This is an ideal society from everybody's microscopic order to the macroscopic order of whole society and nation, and is also a human lofty pursuance. While the society of the rule of law is a type of macroscopic synergetics which possesses macroscopic order, but is not always microscopic order, in which there are various pageant, walkout, etc. A developed direction of the society should be the combination of two aspects, from macroscopic to microscopic, from an actual capable handling to an ideal pursuance.

We proposed the confidence relations and the corresponding influence functions that represent various interacting strengths of different families, cliques and systems in organization.

They can affect products, profit and prices, etc., in an economic system, so the system can produce a multiply connected topological economics [12]. When the changes of the product and the influence are independent one another, they may be a node or saddle point. When the influence function large enough achieves a certain threshold value, it will form a wormhole with loss of capital. It is a mathematical application to economics. This paper is a physical and mathematical application to society.